\documentclass[aps,prd,twocolumn,groupedaddress,showpacs]{revtex4}
\usepackage{amsmath,amssymb,amsfonts,amscd,cancel,graphicx}
\begin{document}

\title{Discrete Noether Currents}
\author{Gerhart  Seidl}
\email[]{gerhart.seidl@gmail.com}
\affiliation{D 95445 Bayreuth, Germany}

\begin{abstract}
A simple implementation of Noether's theorem for discrete symmetries in relativistic continuum field theories is presented. The associated conserved current is exemplified by cyclic symmetries and charge conjugation. In addition, the quantum version of current conservation for discrete symmetries is briefly discussed.
\end{abstract}

\maketitle
\section{Introduction}
Noether's first theorem \cite{Noether:1918zz} expresses probably the most powerful insight into the nature of symmetries in physics. It states that if an action is invariant under a continuous global symmetry transformation, then there exists a conserved current and, thus, a charge that is constant in time. 

This relation has profound implications for all physical systems exhibiting such an invariance. For example, despite having initially been formulated for a classical on-shell theory,  Noether's theorem possesses a quantum analogue in terms of the Ward-Takahashi identities \cite{Ward:1950xp}, which have to be satisfied by any  quantum field theory to ensure renormalizability.

While Noether's original theorem is concerned with continuous transformations, there are also a number of discrete symmetries such as charge-conjugation $C$, parity $P$, time-reversal $T$, and their combinations $CP$ \cite{Christenson:1964fg} and $CPT$ \cite{CPT:1954}, which are of fundamental importance in relativistically invariant theories, such as the standard model. It seems therefore natural to ask how a realization of Noether's theorem for discrete symmetries would look like.

In this paper, we will consider a simple application of Noether's first theorem to discrete groups in usual four-dimensional relativistic field theory. After stating the general formulas in Sec.~\ref{sec:discretesymmetry}, we will calculate explicitly the conserved Noether currents for cyclic symmetries, including charge conjugation, in Sec~\ref{sec:models}. Then, in Sec.~\ref{sec:quantum}, the quantum version of the current conservation for finite symmetries will be commented on. Finally, in Sec.~\ref{sec:conclusions}, a summary and conclusions will be presented.

\section{Currents for Discrete Symmetries}\label{sec:discretesymmetry}
Consider a collection of fields $\Psi_i(x)$, where $i=1,2,\dots$, and $x$ is a spacetime point with components $x^\mu=(t,{\bf x})$, with $\mu=0,1,2,3$ being the spacetime indices. The fields transform under the action of a discrete group $G$ as
\begin{equation}
\Psi_i(x)\rightarrow\widetilde{\Psi}_i(x)=\mathcal{G}^\chi \Psi_i(x),\label{eq:G}
\end{equation}
where $\mathcal{G}$ is a matrix-valued representation of an element of the discrete group $G$ and $\chi \in\{0,1\}$  switches the action of the group on ($\chi =1$)  or off ($\chi =0$).

For (\ref{eq:G}) to be a global symmetry, it must leave -- up to a possible surface term -- the action functional $S[\Psi]$ invariant. It is important that this is achieved without appealing to the equations of motion. In Noether's theorem, however, an essential extra feature is introduced by requiring that the equations of motion -- which are based on infinitesimal variations of the fields -- be satisfied, too. But how do we go about relating the stationary action principle to a symmetry, when it is finite?

One possibility that we will pursue here is to start with field configurations that are invariant under a simultaneous application of the discrete symmetry in (\ref{eq:G}) and a shift of $t$ by some period $T$. That is, the fields shall satisfy an equivalence relation $\Psi_i(t, \vec{x})\sim \widetilde{\Psi}_i(t-T,\vec{x})$. We will then interpret the finite transformation in (\ref{eq:G}) as resulting from a sequence of infinitesimal variations $\chi\cdot\delta \Psi_i(t+\tau,\vec{x})$, where $\tau\in[0,T]$. Each of these variations leads to an infinitesimal change of the action $\delta S[\Psi]_\tau$ that is non-zero, too, since the region of integration of the action is arbitrary. Conversely, due to the discrete symmetry, the integral $\delta S[\Psi]=\int_0^Td\tau\,\delta S[\Psi]_\tau$ will always vanish.

Let us now promote the integer $\chi$ to a position-dependent function $\chi(x)\in\{0,1\}$. Every one of the variations $\delta S[\Psi]_\tau$ will then receive, to leading order in $\chi$, an extra contribution to their respective Lagrangian densities $\propto\partial_\mu\chi(x)$, which is regarded as a weak derivative. Thus, the  change of the action under (\ref{eq:G}), with position-dependent $\chi(x)$, is
\begin{equation}\label{eq:deltaS}
\delta S[\Psi]=\int d^4x\,\mathcal{J}^\mu(x)\partial_\mu \chi(x)\neq 0,
\end{equation}
so that it becomes zero when $\chi(x)$ is constant. Notwithstanding, the contributions to the Lagrangian densities of the $\delta S[\Psi]_\tau$ that are independent from $\partial_\mu\chi$, will, as a result of the discrete symmetry, in total always sum up to zero. If we now assume the equations of motion, the variations $\delta S[\Psi]_\tau$ and, therefore also $\delta S[\Psi]$ in (\ref{eq:deltaS}), will be zero, when $\chi(x)$ vanishes on the boundary. 

Integrating by parts, we see that the current for the discrete symmetry must be conserved, that is,
\begin{equation}\label{eq:divergence}
\partial_\mu \mathcal{J}^\mu(x)=0,
\end{equation}
since the region of integration in (\ref{eq:deltaS}) is arbitrary. As in the continuous case, this implies, for fields that decay at spacial infinity, the existence of a Noether charge
$Q(t)=\int d^3x\,\mathcal{J}^0(x)$
that is a constant of the motion, {\it i.e.},
$\frac{d}{dt}Q(t)=0.$
We will thus call $\mathcal{J}^\mu(x)$ a discrete Noether current for the finite symmetry in (\ref{eq:G}).

Next, let us proceed to obtain an explicit formula for the discrete Noether current. Abbreviating $\Psi_i(\tau)=\Psi_i(t+\tau,\vec{x})$, for $\tau\in [0,T]$, and $\Psi_i(0)=\Psi_i$, we have from the equivalence relation that $\Psi_i(T)=\widetilde{\Psi}_i$. The total change of the action $S[\widetilde{\Psi}]-S[\Psi]$ is then
\begin{equation}
\int_{0}^Td\tau\frac{\delta S[\Psi]_\tau}{\delta \tau}
=  \int_{0}^{T} d\tau \int d^4x\,\dot{\Psi}_i(\tau) \frac{\delta S[\Psi]_\tau}{\delta \Psi_i(\tau)},
\label{eq:path}
\end{equation} 
where, and in what follows, it is understood that boundary terms that arise when evaluating the variational derivatives, are, in general, non-trivial. Since $\widetilde{\Psi}$ is related to $\Psi$ by the symmetry transformation in (\ref{eq:G}), invariance of the action implies that the integrals in (\ref{eq:path}) vanish. On solutions of the Euler-Lagrange equations of motion, we hence find in terms of the Lagrange density $\mathcal{L}(\Psi_i,\partial_\mu\Psi_i)$
that the current
\begin{equation}\label{eq:Jpath}
\mathcal{J}^\mu(x)=
\int_0^Td\tau
\frac{\partial\mathcal{L}}{\partial\big(\partial_\mu \Psi_i(\tau)\big)}\dot{\Psi}_i(\tau)
\end{equation}
is conserved, that is, $\mathcal{J}^\mu$ satisfies (\ref{eq:divergence}), as a consequence of the discrete symmetry. From this, we also see that the charge $Q(t)$ acts as the generator of the discrete symmetry in phase space by means of Poisson brackets as
\begin{equation}
\int_0^Td\tau\Big\{\Psi_i(\tau),\frac{dQ}{dT}\Big|_{T=\tau}\Big\}=(\mathcal{G}^\chi-1)\Psi_i,
\end{equation}
where we have used that $Q(t)$ inherits through (\ref{eq:Jpath}) a dependence on $T$. Similarly, charge conservation can be expressed with the Hamiltonian $H$ as
\begin{equation}
\tilde{H}-H=\int_0^Td\tau\Big\{H,\frac{dQ}{dT}\Big|_{T=\tau}\Big\}=\frac{d}{dt}Q(t)=0,
\end{equation}
where $\tilde{H}$ denotes the Hamiltonian after applying the discrete symmetry and, in the integral, $H$ depends on $\tau$ via the fields and their conjugate momenta.

A description of the discrete Noether current that is equivalent with (\ref{eq:Jpath}) and might be useful in some applications, such as emergent continuous symmetries \cite{Witten:2017hdv}, is based on an expansion in a finite symmetry deformation of $\Psi_i$.
Specifically, we may write (\ref{eq:G}) as
\begin{equation}
\Psi_i(x)\rightarrow\widetilde{\Psi}_i(x)=\Psi_i(x)+\mathcal{F}_i(x),\label{eq:F}
\end{equation}
where $\mathcal{F}_i(x)=(\mathcal{G}^\chi-\mathbf{1})\Psi_i(x)$ is the finite symmetry variation of $\Psi_i$. Invariance of $S[\Psi]$ under (\ref{eq:F}) implies that
\begin{equation}\label{eq:expansion}
\sum_{n=1}^\infty\frac{1}{n!}\Big[\prod_{k=1}^n
\int d^4 x_k\,\mathcal{F}_i(x_k)\frac{\delta}{\delta \Psi_i(x_k)}
\Big] S[\Psi] =0,
\end{equation}
where the variational derivatives act only on $S[\Psi]$ and boundary terms are again non-vanishing. One can then think of (\ref{eq:expansion}) as a Taylor expansion with respect to $\mathcal{F}_i$ and $\partial_\mu\mathcal{F}_i$. Upon using the equations of motion, the $n=1$ term in (\ref{eq:expansion}) becomes $\int d^4x \,\partial_\mu\mathcal{J}^\mu_\text{a}(x)$, where $\mathcal{J}^\mu_\text{a}=\mathcal{F}_i\partial\mathcal{L}/\partial\partial_\mu\Psi_i$ may be interpreted as the Noether current of an emergent continuous symmetry.

The conserved Noether current $\mathcal{J}^\mu(x)$ for the discrete symmetry then reads
\begin{equation}\label{eq:J}
\mathcal{J}^\mu(x)=\mathcal{J}^\mu_\text{a}(x)+\mathcal{J}^\mu_\text{b}(x),
\end{equation}
where
\begin{equation}
\mathcal{J}^\mu_\text{b}(x)=\frac{i}{(2\pi)^4}\int d^4 k\, d^4y\,\frac{k^\mu\,\mathcal{N}(y)}{(k_0+\text{i}\epsilon)-{\bf k}^2}e^{-ik(x-y)},\label{eq:Jb}
\end{equation}
and $\mathcal{N}$ is defined as the sum in (\ref{eq:expansion}) starting with $n=2$. In the context of an emerging continuous symmetry, $\mathcal{J}^\mu_\text{b}$ would then contain the irrelevant operators that break the continuous symmetry to the discrete subgroup $G$.

\section{Simple Examples}\label{sec:models}
We will now apply our results to a class of simple internal cyclic symmetries  ({\it cf.} \cite{Krauss:1988zc}), which have, {\it e.g.}, been used to address the problem of fermion masses and mixings (for a systematic scan, see \cite{Plentinger:2008nv}). In particular, we will relate the simplest case to charge conjugation $C$, the breaking of which is believed to be crucial for understanding the matter-asymmetry of the universe \cite{{Sakharov:1967dj}}.

\underline{\it Cyclic symmetries:} Let us consider a complex scalar field $\phi(x)$ with Lagrangian
\begin{equation}
\mathcal{L}(x)=|\partial_\mu\phi|^2+\lambda(|\phi|^n+\text{h.c}),\label{eq:phin}
\end{equation}
where $n$ is an integer and $\lambda$ is a real coupling. The Lagrangian exhibits a global $Z_n$-symmetry $\phi\rightarrow e^{\text{i}2\pi/n}\phi$.

To extract the relevant degree of freedom, we write $\phi(x)=v\cdot e^{\text{i}\theta(x)}$, where $v$ is the vacuum expectation value of $\phi$ and $\theta$ is a real Nambu-Goldstone-type field. For simplicity, we will assume that $\theta$ is spatially constant.

To first oder in $\lambda$, the equation of motion is solved by
\begin{equation}
\theta(t)=\theta_0(t)+2\frac{\overline{n}\lambda v^n}{\omega^2}\text{sin}\big(\overline{n}\,\theta_0(t)\big)+\mathcal{O}(\lambda^2),
\end{equation}
where $\overline{n}=\frac{n}{\sqrt{2}v}$, $\omega=\frac{2 \pi}{T}$, and $\theta_0(t)$ is the zero mode
\begin{equation}
\theta_0(t)=\theta_0+\frac{\omega t}{\overline{n}}\quad\text{mod}\bigg(\frac{2 \pi n}{\overline{n}}\bigg)\label{eq:zero},
\end{equation}
where $\theta_0=\theta(0)$ and we have assumed the equivalence relation $\theta(t)\sim\theta(t-T)$.
The $Z_n$-symmetry acts on $\theta_0$ and $\theta$ via $t \rightarrow  t+T$ as a shift-symmetry $\theta\rightarrow\theta+2\pi/\overline{n}$.
In terms of the canonical momentum $\pi_0=\dot{\theta}(0)$,
the Noether current (\ref{eq:Jpath}) for the $Z_n$-symmetry is
\begin{equation}
\mathcal{J}^0=\frac{2\pi}{\overline{n}}\pi_0\left(1-2\lambda\frac{v^n}{\pi_0^2}\right)+\mathcal{O}(\lambda^2),\label{eq:JZn}
\end{equation}
which is time-independent and, thus, conserved. The discrete Noether currents for models that are based, for instance, on non-Abelian internal symmetries, such as $A_4$ \cite{Babu:2002dz}, can be determined similarly.

\underline{\it Charge conjugation:} If $n=2$, besides the additive shift symmetry $\theta\rightarrow\theta+\pi\sqrt{2} v$, the field $\theta$ possesses an equivalent representation $\theta(t+T)=\pm\theta(t)$ of the $Z_2$ symmetry. We can identify the eigenvalues of this transformation with parity $\eta_C$ under charge conjugation $C:\theta\rightarrow\eta_C\,\theta$, where $\eta_C=\pm 1$, for $\theta$ being even/odd under $C$. The zero mode in (\ref{eq:zero}) then becomes $\eta_C^{\lfloor t/T \rfloor}\big(\theta_0(t)\;\text{mod}(\pi n/\overline{n})\big)$ and (\ref{eq:JZn}), for $n=2$, yields the conserved current for $C$.
 
\section{Comment on the Quantum version }\label{sec:quantum}
The current divergence $\partial_\mu\mathcal{J}^\mu$ of a classically conserved Noether current $\mathcal{J}^\mu(x)$ will appear in the quantum theory as a  symmetry generator localized at $x$. At the quantum level, the discrete Noether current will then not only be conserved but also exhibit Schwinger-Dyson equations associated with the current conservation that are similar to the case of the global symmetry being continuous.

To illustrate, consider for a scalar field $\phi(x)$ a discrete symmetry $\tilde{\phi}=\mathcal{G}^{\chi}\phi$ as in (\ref{eq:G}), with a classically conserved discrete Noether current $\mathcal{J}^\mu$. The path-integral describing the $2$-point function $\langle\phi(x_1)\phi^\ast(x_2)\rangle$ is then invariant under the unitary transformation $\tilde{\phi}(x)=\mathcal{G}^{\chi(x)}\phi(x)$, which need not be a symmetry of the action.

The correlator of $\partial_\mu\mathcal{J}^\mu$ with $\phi(x_1)$ and $\phi^\ast(x_2)$ then satisfies a Ward-Takahashi identity analogous to that for a continuous symmetry with contact terms
$ \text{i}\langle\text{ln}\,\mathcal{G}\phi(x_1)\delta(x-x_1)\phi^\ast(x_2)\rangle+\dots$ In the same way, we arrive at the Schwinger-Dyson equations and analogues of the Ward-Takahashi identity for discrete symmetries for the other $n$-point correlation functions.

\section{Conclusions}\label{sec:conclusions}
Albeit Noether's first theorem has originally been formulated for continuous symmetries, it is possible to extend it also to discrete transformations, which play an important role in any Lorentz-invariant field theory.

The discrete version of Noether's theorem considered here refers to genuinely discrete symmetries in usual continuum field theories. The conserved current for discrete symmetries can be constructed (i) by tying together a sequence of infinitesimal shifts in the fields that build up the discrete symmetry transformation or (ii) by expanding the discrete variation of the Lagrangian in finite field deformations. Just as it is true for continuous invariances, the associated conserved discrete charge acts as a generator of the discrete symmetry, too. As a showcase, it has been exhibited how to apply this to find the discrete Noether currents in simple examples. Furthermore, it has been highlighted that discrete Noether currents yield Schwinger-Dyson equations or Ward-Takahashi identities analogous to the case of continuous global symmetries. The implementation of the discrete Noether's theorem for realistic field theories incorporating, {\it e.g.}, the gauge group and fields of the standard model is straightforward.

Also in view of the vast amount of possible finite groups \cite{Gorenstein:1983}, it is conceivable that these results may offer tools for theories that utilize discrete symmetries to describe the observed patterns of parameters in Nature or other phenomena, such as new states of matter \cite{Wilczek:2012jt}.

\section*{Acknowledgements}
I would like to thank R.~R\"uckl for pointing out the restriction of Noether's first theorem to continuous groups.

\end{document}